# Methane detection to 1 ppm using machine learning analysis of atmospheric pressure plasma optical emission spectra.

Tahereh Shah Mansouri[1]*, Hui Wang[2], Davide Mariotti[1], Paul Maguire[1]

[1]NIBEC Engineering, Ulster University and [2]Queens University, Belfast, N. Ireland.

* Corresponding author:

## Abstract

Optical emission spectroscopy from a small-volume, 5 µL, atmospheric pressure RF-driven helium plasma was used in conjunction with Partial Least Squares – Discriminant Analysis (PLS-DA) for the detection of trace concentrations of methane gas. A limit of detection of 1 ppm was obtained and sample concentrations up to 100 ppm $CH_4$ were classified using a nine-category model. A range of algorithm enhancements were investigated including regularization, simple data segmentation and subset selection, feature selection via Variable Importance in Projection and wavelength variable compression in order to address the high dimensionality and collinearity of spectral emission data. These approaches showed the potential for significant reduction in the number of wavelength variables and the spectral resolution/bandwidth. Wavelength variable compression exhibited reliable predictive performance, with accuracy values > 97%, under more challenging multi-session train – test scenarios. Simple modelling of plasma electron energy distribution functions highlights the complex cross-sensitivities between the target methane, its dissociation products and atmospheric impurities and their impact on excitation and emission.

1. Introduction

Gas identification and in particular the detection of trace levels of molecular components in gases has gained increasing attention in many fields from atmospheric pollution and climate change monitoring to industrial safety [1] and breath analysis for clinical diagnosis.[2] There are a number of established techniques including mass spectrometry[3], gas chromatography[4] optical spectroscopy, electrochemical,[5] solid-state and optic fibre,[6] that have inspired the development of a wide range of technologies in each category. Laser absorption and spectroscopic detection methods such as non-dispersive IR absorption (NDIR) or Raman have allowed limits of detection (LOD) in the low ppm to ppb range to be achieved. Improvements in, for example, mid-IR quantum cascade laser technology and photoacoustic detectors will enable continued reduction of LOD. Among the spectroscopic techniques, tuneable diode laser IR absorption spectroscopy (TLDAS) and atomic emission spectroscopy ICP-AES are well-established and routine laboratory techniques. Apart from improving LOD and increasing the number of target species, there is a major drive towards system miniaturisation and cost reduction in order to achieve field deployable gas detection capability e.g. for rapid and continuous environmental monitoring via autonomous distributed networks or point of care clinical breath screening. For example, methane is a high priority greenhouse gas with stringent targets for reduction, including reducing $CH_4$ emissions from e.g. landfill, oil and natural gas industries.[7-10] Field deployment of high-resolution detectors and remote autonomous monitoring is a major priority yet remains elusive due to the very high system cost.[10] This has inspired the search for high accuracy miniaturised systems. The ARPA-E (US) MONITOR programme has funded development of various technology strands including compact IR spectrometry, compact mass spectrometry, hollow-core optic fibre and low cost printed nanomaterials.[11] Methane is also an important breath biomarker and detection of trace $CH_4$ levels is a major challenge. Recently Dong et al. reported a compact trace $CH_4$ detection system based on TLDAS with distributed feedback interband cascade lasers in a 5 litre volume package.[12] However, these compact systems remain costly when considered for autonomous field deployment. Detector arrays based on high porosity and high surface area nanomaterials have been proposed as a low-cost electronic nose platform for breath and environmental analysis. High sensitivity has been achieved when coupled with machine learning [13] but systems struggle with lifetime, species interference and cross-sensitivity (e.g. temperature and humidity).[14,15] The use of plasmas as atomic and molecular sources for optical emission spectroscopy (OES) and mass spectrometry (MS) has a long history, with the ICP-AES technique being the most popular. Samples in the form of liquid or solid particles are introduced into the hot plasma resulting in vaporisation or evaporation, excitation and ionisation and provide either the photons for OES or ions for MS.[16] Low pressure glow discharge plasmas and laser induced breakdown spectroscopy (LIBS) are typically employed to produce OES species from solid surfaces.

Recent progress in the design and control of miniaturised atmospheric pressure plasmas systems has encouraged their application to new research fields such as e.g. plasma-based medicine, agriculture, gas reforming, catalysis, advanced nanomaterials and environmental pollution control.[16,18,19] The associated plasma devices are of simple construction, small, low cost and operate at atmospheric pressure and at low temperatures. They can also provide high intensity light emission and therefore have the potential to act as an optical emission source for trace gas detection. Hyland et al. first reported plasma OES with machine learning for trace gas detection and recognition, using spectra from a range of trace volatiles fed into neural networks. [17] Weagant et al. investigated the use of a low power atmospheric pressure Ar – $H_2$ microplasma and portable spectrometer to detect trace metal impurities. Liquid samples were dried then electrothermally vaporised into the plasma. Simple spectra resulted, dominated by low excitation energy lines. However maintaining reproducible line intensities and limiting background emission was difficult.[20] Similar trace metal detection in liquid has been demonstrated using RF-excited glow discharge emission spectroscopy at atmospheric pressure.[21] Here the sample is dried and then ablated by a plasma operating at gas temperature up to 1500 °C. The observed spectral variability was up to 28%. Atmospheric pressure plasmas in contact with water and complex liquids have been investigated for rapid, lower cost and low power analytical atomic spectrometry of metals.[22-26] High plasma densities (> $10^{20}$ m$^{-3}$) [27] and relatively high gas temperatures (600 K – 1100 K) are involved [27,28] and mechanisms depend on liquid evaporation and droplet formation while the inclusion of organic species can enhance emission or produce volatile species containing the elements for

detection.[29]

While trace metal detection by miniaturised plasmas may offer low cost portable alternatives to ICP-AES, the trace gas detection of molecular and complex volatile constituents represents a much greater challenge since microplasma emission spectra are very complex, individual lines are weak and poorly resolved, especially for non-equilibrium low temperature (NELT) plasma devices. High resolution OES of NELT plasmas containing molecular mixtures is often used to fit observed to simulated spectra in order to determine internal plasma parameters such as gas rotational and vibrational temperatures [30] as well as electron temperature and density.[31] We have carried out such analysis on similar plasma devices to that used here.[32,33] However with objectives such as portability, low cost, field deployability and possible autonomous operation, the intrinsic complexity of the spectra and the temporal variation in plasma conditions under uncontrolled conditions need to be considered. Using design constraints and operating parameters that maintain low gas temperatures (< 50 $^{o}$C), e.g. for breath analysis or managing safety concerns with flammable gases, adds further noise and complexity to spectra. Knowledge of NELT plasma chemistry is very limited and generating accurate simulated spectra for molecular gases and mixtures, especially at trace concentrations below 100 ppm, is not feasible. This coupled with the use of low-cost limited resolution spectrometers presents a major impediment to accurate detection and to date, the use of OES with NELT plasmas to determine the trace molecular constituents of a gas has not been considered. Kudryavtsev et al. used a current probe technique integrated into a helium microplasma, for $CO_2$ gas analysis via collisional electron spectroscopy. This involved measurement of the high energy portion of the Electron Energy Distribution Function (EEDF) to determine He metastable reactions with impurities. $CO_2$ detection at concentrations ≥ 500 ppm was achieved. However the plasma was operated below atmospheric pressure.[34] Recently, we demonstrated the feasibility of using OES to detect the presence of methane above threshold values in the low ppm range using a helium NELT plasma jet coupled with spectral analysis via machine learning techniques.[35]

Spectral data is often used to help determine the constituents of materials and can consist of, for example,

measured values of radiation or mass intensity at fixed discrete wavelengths or mass values, respectively. Chemometric and machine learning techniques have been applied where spectral discrimination is problematic.[36] The interpretation of optical spectra, from UV to far IR, depends on the experimental approach and the instrument resolution. Thus with absorption spectra, the concentration of a target species may be directly related to measured intensity through the Beer-Lambert Law. For emission spectra, a relationship between concentration and intensity, at a specific wavelength, is only possible when the system is in local thermodynamic equilibrium, in which case it is determined from the Boltzmann distribution which relates excited state densities to that of the ground state. High temperatures are therefore required to obtain measurable emission intensities. For the low temperature emission spectra used here, thermodynamic equilibrium is not established and spectral data consist of a large number of lines where there is little a priori information about expected line strength and significance. Intensity values will follow a complex non-linear relationship with concentration. Line broadening via intrinsic or instrumental effects will create data values around each peak which may be highly correlated and redundant and/or merge peaks from different excitation states and species. Important molecular gases, including hydrocarbons such as methane, generally have multiple but weak lines in the UV-Vis-NIR region which often overlap with spectral lines from plasma carrier gases, such as helium or argon, and impurities. Furthermore, the introduction of molecular gases into a plasma can affect parameters such as electron density and temperature which in turn modify line intensities of atomic and impurity gases (e.g. $O_2$, $N_2$ and $H_2O$ dissociation products).

In order to cope with such complexity, we focus on developing machine learning algorithms for analysis of NELT plasma emission spectra which can handle the challenges of high dimensionality, where the number of variables (wavelengths) greatly outnumbers the sample count, nonlinearity, redundancy, collinearity, where individual peaks bleed into multiple nearby data points, and multimodality. [37] Recently, we developed a number of algorithmic approaches based on PLS-DA to characterise reflectance spectral data from portable optical and infra-red systems under uncontrolled and variable field conditions. [37-41] Algorithm performance was also compared with traditional laboratory-based absorption spectra. Emission spectra, by contrast, display a much larger number of sharp well-defined peaks with a wide range of intensities and thus the algorithmic challenges are heightened. Recently, machine learning approaches have been investigated in an attempt to solve critical unresolved challenges in real-time diagnostics and control of cold atmospheric pressure plasmas. These include monitoring vibrational/rotational temperatures and the effects of changing substrate properties on plasma conditions, [42,43] and to determine the electron energy probability function solely from optical emission spectra. [44] Using a coplanar high-voltage AC plasma and spectral analysis based on convolutional neural networks, Wang et al. demonstrated detection of methanol and acetone in real-time for concentrations above 1487 ppm and 3439 ppm respectively.[45] In this work, we seek to extend our initial feasibility study [35] to identify impurity species at different and lower concentrations using multi-categorical models. Emission spectra from methane in helium mixtures, with concentrations from 0 to 100 ppm, were obtained from a low cost portable NELT plasma device. The gas mixtures also contained trace impurities from air and $H_2O$ of unknown concentration. Machine learning models based on PLS-DA were investigated, using a range of training and test protocols, along with a number of data manipulation and feature selection approaches in order to maximise performance.

## 2. Experimental Methods

$CH_4$ – He spectra were obtained from an RF-excited (13.56 MHz) plasma formed in a quartz capillary between two exterior ring electrodes (separation 5 mm) while helium was used to sustain the plasma, figure 1. The 0.7 mm (ID) capillary outlet was a large distance (~100 cm) from the plasma to minimise atmospheric impurity back-diffusion. The system was initially conditioned to remove background impurities from the capillary walls, using repeated daily exposure to a 100% argon plasma, over 21 days, followed by isolation and continuous exterior IR heating of the capillary. A two-stage mass flow-controlled gas network was used to dilute methane gas (purity 99.95%) in the carrier gas (He, purity 99.9995%). Two mass flow controllers (MKS, Model 1179C, precision 0.05% FSD) were used to deliver up to 0.005 SLM of He – $CH_4$ mixture at a concentration of 100 ppm into a pure He flow up to 0.05 SLM. The overall specified precision at 1 ppm $CH_4$ was ±1%. The set concentration error included manufacturer specified flow meter error (± 1% FS) and gas supplier (Buse International) specified $CH_4$ in He mixing accuracy (± 5%). The maximum deviation from set concentration was + 30.9% / - 26.8% and the absolute deviation values are shown in

Figure 1 (b). The gas temperature, measured in a similar plasma system, remained below 30 °C.[46]

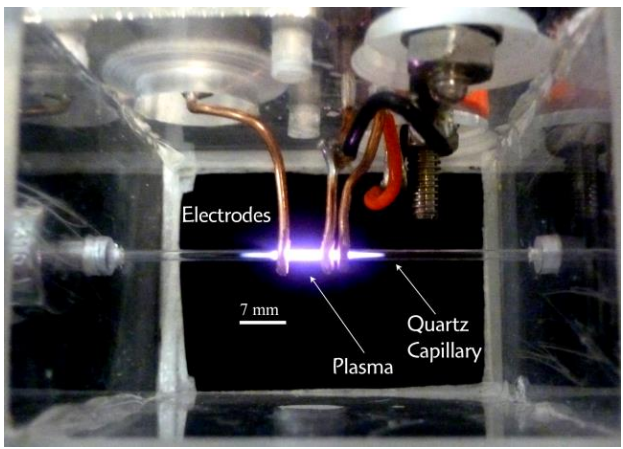

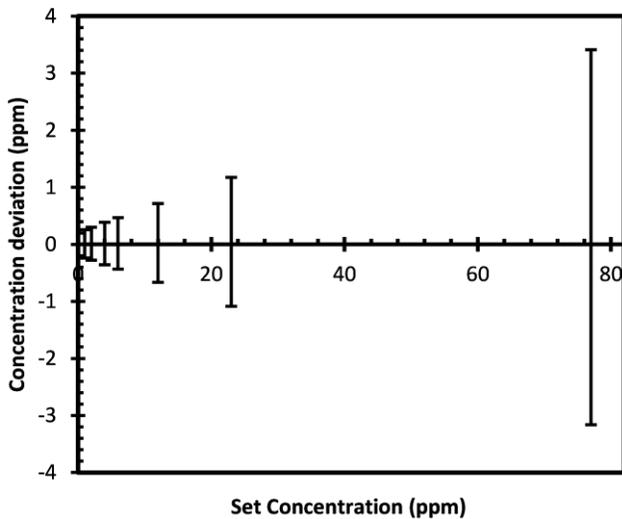

Figure 1: (a) NIBEC RF capillary plasma system operated with He carrier gas at atmospheric pressure. The electrode gap was 5 mm and the internal diameter of the capillary was 0.7 mm, (b) Set concentration deviation due to flow meter specified error and supplied gas mixing accuracy.

$CH_4$ – He data was collected in two separate datasets, where dataset A comprised 523 samples in nine $CH_4$ concentration categories (0, 1, 2, 4, 6, 12, 23, 77, 100 ppm) and dataset B comprises 720 samples in eight (0, 1, 2, 4, 6, 23, 77, 100 ppm) $CH_4$ concentration categories. Spectra in the wavelength range 194 nm – 1122 nm (interval 0.25 nm) were obtained using an Ocean Optics HR4000CG-UV-NIR spectrometer (optical resolution < 1.0 nm FWHM, slit width 5 µm), with a total of 3648 wavelength points recorded. Spectral mean intensity versus wavelength from 0 ppm, 2 ppm and 100 ppm $CH_4$ sample sets are shown in Figure 2 (a). Spectral change with concentration is indicated in difference plots, Figure 2 (b) while intensity change between samples is indicated by the relative standard deviation at each wavelength, Figure 2 (c). The main spectral lines are listed in Table I in rank order of intensity at 0 ppm $CH_4$ and intensity values relative to the intensity of the largest peak at 588 nm are indicated. Using spectral intensity data, species involved in specific transitions are listed.[47,48] Impurity lines, representing species derived from air (N2, O2, N, O) and water dissociation (OH, H), are noticeable with intensities up to 20% of the maximum. The $C_2$ (Swan) vibrational bands around 516 nm are only visible at concentrations ≥ 77 ppm. The only other detectable lines that may be attributed to $CH_4$ fragmentation are the CH(A – X) band at 388.90 nm which overlaps with the He line at 388.86 nm and possibly $N_2$ lines at 389.46 nm. The integrated line intensity taken over the range 388 nm ±3 nm exhibits an approximately constant value at low concentrations, suggesting that CH (A – X) emission may not be significant until 77 ppm, where the intensity is noticeably enhanced.[35] The $H_\alpha$ line intensity at 656.56 nm, which may derive from $H_2O$ dissociation and/or $CH_4$ fragmentation, varied approximately linearly with $CH_4$ concentration and at ≥ 77 ppm was greater than that of the main He line at 588 nm.[35]

| Rank | Peak wavelength (nm) | Relative Intensity | Species |
|---|---|---|---|
| 1 | 588 | 1.00 | He |
| 2 | 706 | 0.79 | He |
| 3 | 667 | 0.21 | Impurity, He |
| 4 | 778 | 0.10 | Impurity |
| 5 | 389 | 0.08 | He, CN, $N_2$, $O_2$ |
| 6 | 336 | 0.05 | Impurity |
| 7 | 728 | 0.04 | He, impurity |
| 8 | 656 | 0.03 | H |
| 9 | 415 | 0.02 | He I |
| Selected peaks of intensity rank > 9 or which only appear for $CH_4$ ≥ 77 ppm | | | |
| | 516 | - | C2 Swan |
| | 309 | - | OH |
| | 431 | - | CH |

**Table 1 Main OES peaks of $CH_4$ - He listed in rank order of intensity as observed for 0 ppm $CH_4$ except for features around 516 nm which are only observed at ≥ 77 ppm $CH_4$. The peak wavelengths have been rounded to nearest integer values. The relative intensity column values are calculated with respect to the maximum peak intensity (588 nm) at 0 ppm $CH_4$. The species column lists the attributed species and for those wavelengths with overlapping species peaks, the species are listed in order of expected intensity.**

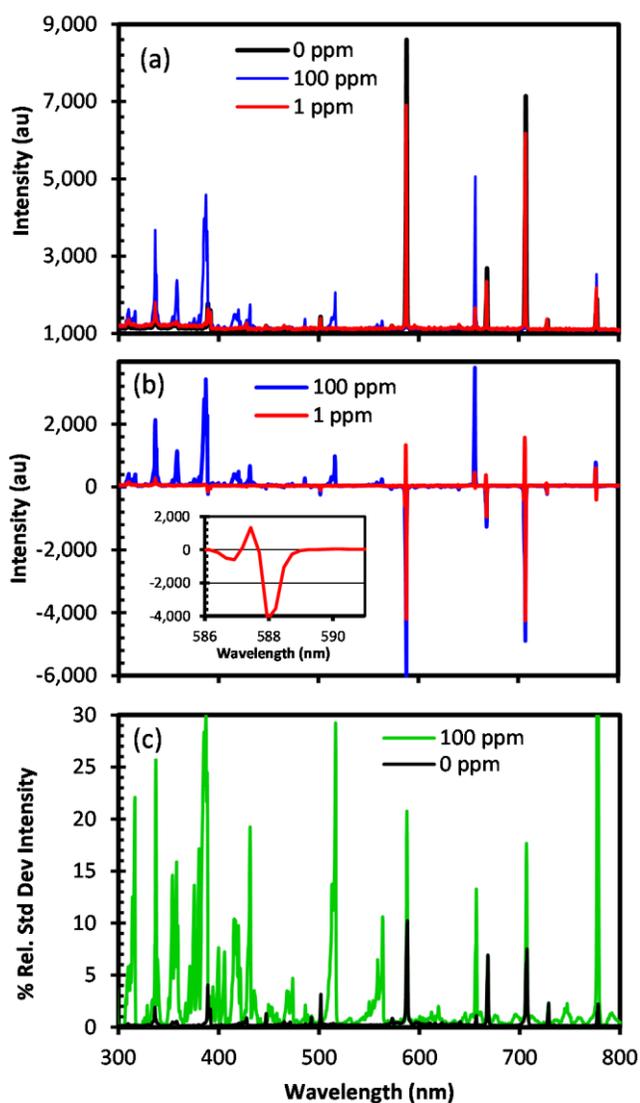

Figure 2: (a) Mean spectra from samples with 0 ppm, 1 ppm and 100 ppm CH$_4$, truncated to the wavelength range 300 nm – 800 nm, (b) Difference plots of 1 ppm CH$_4$ - 0 ppm and 100 ppm CH$_4$ - 0 ppm. The inset shows the difference around 588 nm for 1 ppm CH$_4$ and indicates the effect of misalignment as a contributor to the observed difference. (c) % relative standard deviation (SD/mean) for 0 ppm and 100 ppm spectra.

## 3. Computational Methods

The overall objective is to develop an algorithmic solution to the task of recognising an unknown spectrum as a member of one category. In an exploratory search, the raw data was subjected to various pre-processing steps. These included Standard Normal Variation (SNV), normalization, baseline correction, auto scaling and noise reduction. Initially, we looked briefly at the performance profile of four different algorithmic approaches (PLS-DA, KNN, SVM-PCA, LDA) using the Receiver Operating Characteristic (ROC) curve, Figure 3, for a single category (0 ppm CH$_4$). As we previously found with infra-red spectra

[37,39] the PLS-DA algorithm shows the best ability to distinguish spectral data from any two groups. For example, in this case distinguishing 0 ppm $CH_4$ concentration from those ≥ 1 ppm $CH_4$, the area under curve (AUC) was > 98% for PLS-DA while Linear Discriminant Analysis (LDA) [49] showed the poorest classification at 64%. Both weighted K-Nearest Neighbour (wKNN) [50] and Support Vector Machine coupled with Principal Component Analysis (SVM-PCA) [51,52] show intermediate performance.

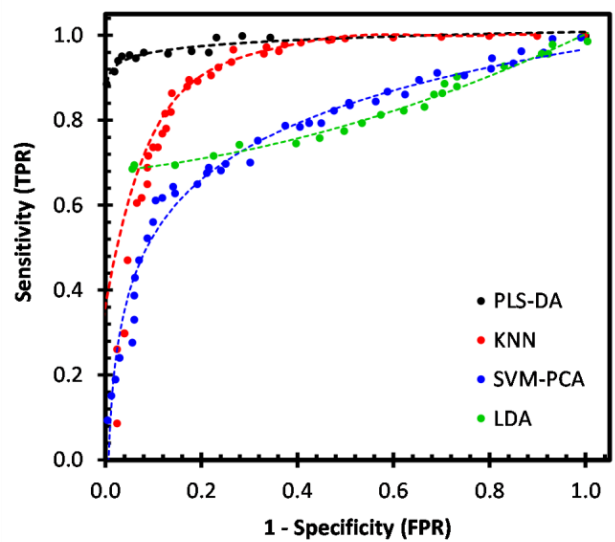

**Figure 3: Comparison of Receiver Operating Curves (ROC) for four algorithms (PLS-DA, wKNN, SVM-PCA and LDA) applied to pure He spectra.**

Partial least squares discriminant analysis (PLS-DA) is a classification derivative of PLS regression and is considered a useful algorithm for building predictive models in cases where there is both a large number of parameters and factors which are highly collinear and has been used regularly in the analysis of chemometric data. A Variance Inflation Factor (VIF) greater than 10 indicates harmful data collinearity and a reason for concern and almost all $CH_4$ – He spectra data display VIF values >10. [53-56] PLS models the relationship between an input matrix (X) and an output matrix (Y), to develop an N-dimensional hyperplane in the input X space that is related as closely possible to the output response matrix Y. PLS-DA searches for linear combinations of independent (predictor) variables, namely latent variables (LV), that maximize the covariance between the latent variable and the response. Furthermore, where the Y data measurements can be classified into different independent categories, i.e. trace gas concentrations, the algorithm is capable of setting separate and simpler models for each Y category.

PLS-DA is implemented using the SIMPLS algorithm in Matlab.[57]

| Protocol | Description |
|---|---|
| 1 | Model training and cross-validation using dataset A |
| 2 | Model training and cross-validation using dataset B |
| 3 | Datasets A and B merged. Model training and cross-validation using merged A + B data |
| 4 | Model trained and cross-validated using dataset A. Model testing using dataset B |

Table 2: Protocols for implementation of algorithm training, evaluation and testing

Models were constructed using nine output categories for dataset A and eight for dataset B. Initially, model performance was evaluated using 4 different protocols as listed in Table 2. Protocols 1 – 3 represent within individual or combined session evaluations while Protocol 4 uses one dataset for training and the other for testing. In each protocol, the ratio between training and validation samples is 50-50%. With cross validation, different subsets of the data are used for training and testing and the accuracy of model prediction with unseen test data is determined. This procedure is repeated with different data subsets to provide an estimate of average prediction accuracy and the Root Mean Square Error (RMSE). [58]. The Leave One Out Cross Validation (LOO-CV) procedure uses all samples but one as the training set, the remaining sample acting as the blind test. This procedure is repeated until all samples are used as test and the mean accuracy and RMSE are returned. The PLS-DA algorithm was tested using a model set, where each individual model was constructed using 1 to 15 latent variables. Leave One Out Cross Validation (LOO-CV) approach was applied to each model to acquire an estimate of the model accuracy versus the number of latent variables (LV) used to build the model. [58] We investigated a number of enhancements to the PLS-DA analysis, detailed below, in order to evaluate and improve algorithm prediction accuracy.

### 3.1. Regularisation

Model overfitting, whereby the high accuracy obtained in model training is not replicated in the test phase can often be reduced using standard regularisation techniques whereby a penalty term is introduced to constrain some of the model regression coefficients.[59-61] Three common algorithms were investigated, namely Lasso, Ridge and Elastic Net. Lasso regularisation (L1 norm) forces the sum of the absolute value of the regression coefficients to be less than a fixed value which in turn forces some coefficients to zero, removing them from the model and its penalty term can be written as $\lambda \sum_{j=1}^{p} |\beta_j|$ where $\lambda$ is the regularization parameter that determines how much the model's flexibility should be penalized and $\beta_j$ is the regression coefficient. In contrast, the penalty term for Ridge regularisation (L2 norm) can be defined as $\lambda \sum_{j=1}^{p} \beta_j^2$ resulting in all coefficients being regularised equally but with a much smaller number of coefficients set to zero. Elastic net creates a linear combination of the L1 and L2 regularisation penalties by adding a quadratic, i.e. Ridge, penalty to that of Lasso with a constant $\alpha$ determining the relative weights and is given by $\lambda(\frac{(1-\alpha)}{2}\sum_{j=1}^{p}\beta_j^2 + \alpha \sum_{j=1}^{p}|\beta_j|)$.[62]

### 3.2. Data segmentation

To investigate the impact of data redundancy and the large number of data variables on model prediction, each original dataset of 3648 variables (wavelengths) was split into M subsets, each containing N variables. For a given M, N and LV, models were then built for each data subset and accuracy compared. This was an exploratory task with the objective of providing qualitative insight into how spectral characteristics may affect predictions and hence a systematic variation of M, N and LV was not carried out. It is an informal approach which compares different individual subset models unlike interval PLS (IPLS) which performs an exhaustive search and then adds subsets sequentially to the model.

### 3.3. VIP Selection

The relative importance of each input variable in modelling the output response can be determined from the Variable Importance in Projection (VIP) scores. These measure the contribution to the model of each predictor variable, j, by accounting for the covariance between $X_i$ and $y_i$, where i is the $i^{th}$ latent variable, as expressed by the calculated PLS weights $(W_{i,j})^2$ in (1), [63,64]

$$\text{VIP}_j = \sqrt{\frac{\sum_i^n S^2(y,t_i)(\frac{w_{ij}}{w_i})^2}{(\frac{1}{m})\sum_i^n S^2(y,t_i)}} \qquad (1)$$

where $m$ is the total number of predictor variables, $n$ is the total number of latent variables and $s^2(y,t_i)$ is fraction of y variance defined by latent variable $i$. Subsequent revised models can then be generated using a reduced set of input variables whose VIP scores are above a given threshold. A common approach assumes a threshold greater than 1, which is the average of the squared VIP scores, thereby selecting variables with an above average contribution to the model.

### 3.4. Peak width compression

From [35] we have found that data in regions around spectral peaks makes the most important contribution to the simplified binary classification. Also the model accuracy was found to be very sensitive, in some conditions, to peak measurement misalignment due to spectrometer jitter. This misalignment caused peaks in similar samples to appear up to a few variable units away from its nominally true value and is interpreted by the model as separate variables. To counter this, the variable values of a number of major peaks were established as references and spectra subjected to alignment shifting. However the required shift was non-linear and the existence of reliable reference peaks below 500 nm could not be guaranteed for all conditions. Therefore an alternative approach was investigated. Where the underlying optical transition is expected to be a line transition at a single wavelength subject to instrumental broadening, each measured peak spans a range of wavelengths due to the low resolution and inherent jitter of the spectrometer. Therefore in this approach the observed broad peak, over a wavelength range $\Delta\lambda$, is compressed to a single intensity value by summing the intensities over $\Delta\lambda$. This value is then assigned to a single wavelength variable. The remaining variables within $\Delta\lambda$ are then discarded from the model. This procedure is carried out on each peak of intensity greater than a set threshold (100) for each category using the Savitzky-Golay (SG) algorithm for data smoothing and peak finding. This method aims to remove correlated variables around a peak while assigning their summed intensity to the single peak variable and also avoiding the issue of misalignment.

## 4. Results

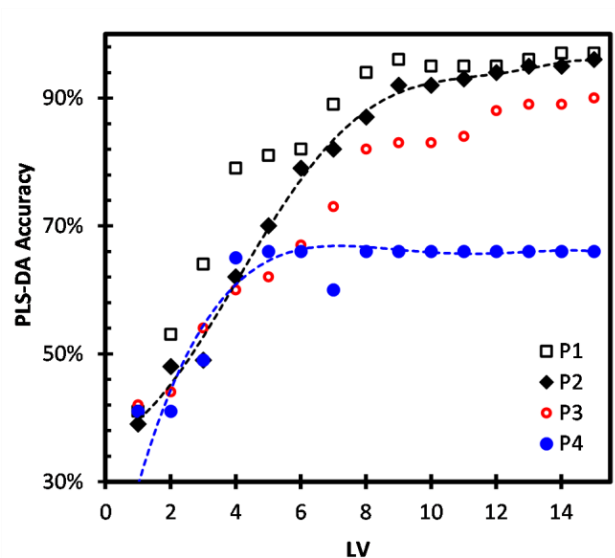

**Figure 4: Comparison of PLS-DA accuracy versus the number of model latent variables for four protocols given in Table 2. Error bars, representing the RMSE values at each latent variable, are of similar size to symbols for LV > 5 and are excluded for clarity.**

Multi-category PLS-DA models were constructed from He – $CH_4$ spectra with varying $CH_4$ concentrations up to 100 ppm, using the full 3668 wavelength variable set over a range of LV values for each of the protocols in Table 2. Using cross-validation (CV) the model accuracy range for each LV was determined and results are given in Figure 4. As expected, accuracy improves with LV and tends to saturate for LV > 8. While the outcomes for protocols 1 and 2 are similar, the accuracy is lower when both datasets are merged, protocol 3. With protocol 4 the test data is obtained from a different session to that of the training data and the outcome is a poor classification accuracy across all LV values. With the addition of regularisation, the penalty factor, $\lambda$, was determined from 10-fold CV of the LASSO algorithm. The variation in mean square error (MSE) with $\lambda$ is given in Figure 5 with $\lambda_{min} = 10^{-2.25}$ at the lowest MSE. However implementing either LASSO or Ridge regularisation with $\lambda$ from $\lambda_{min}$ to $10^{-2}$, where the sparsest models are formed, resulted in limited improvement in accuracy. The LASSO regularisation identified a subset of 275 wavelength variables, from the original ~3600, for use in the model. While this indicates a high degree of variable redundancy, over 30% of the selected variables were from the long wavelength range (> 800 nm) which is featureless and highlights the likelihood of noise amplification as a by-product of the penalty term.

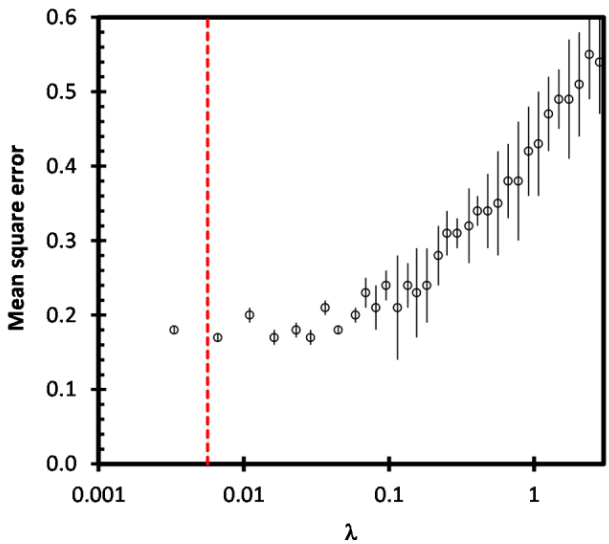

Figure 5: Mean square error (MSE) versus the LASSO penalty $\lambda$ from 10-fold CV. The minimum cross-validated MSE occurs at $\lambda = 10^{-2.25}$ (red line) while the sparsest model with low MSE occurs at $\lambda = 10^{-2}$.

An exploration of the regions of the spectra important to model accuracy was undertaken via data segmentation into M subsets, each containing N wavelength variables. M subset models were constructed using protocol 2 and an LV value of 15. In Figure 6 the accuracy is compared for each subset model for both training and test data, where M is 36 and N is ~100. Overall the outcome shows a degree of overfitting that is most pronounced in the featureless regions at long and short wavelengths, while the lowest degree of overfitting occurs in the wavelength range 300.62 nm – 327.10 nm. The best subset model accuracy was observed for subsets in the wavelength range 511.02 nm – 536.73 nm with accuracy 92% similar to that achieved from full variable models (95%). The wavelength ranges of the top 5 test model accuracies correlate well with the highest spectral peaks, Table 3.

| Subset No. | Wavelength Interval (nm) | Subset Accuracy % | Peak Intensity rank |
|---|---|---|---|
| 15 | 562.64 – 588.47 | 91.33 | 1 |
| 13 | 511.02 – 536.73 | 92.58 | 2 |
| 5  | 300.62 – 327.10 | 92.49 | 3 |
| 11 | 458.86 – 484.71 | 91.89 | 4 |
| 16 | 588.47 – 614.23 | 90.15 | 5 |

Table 3: Wavelength ranges of the top 5 test model accuracies and their relation to the spectral peak height ranking.

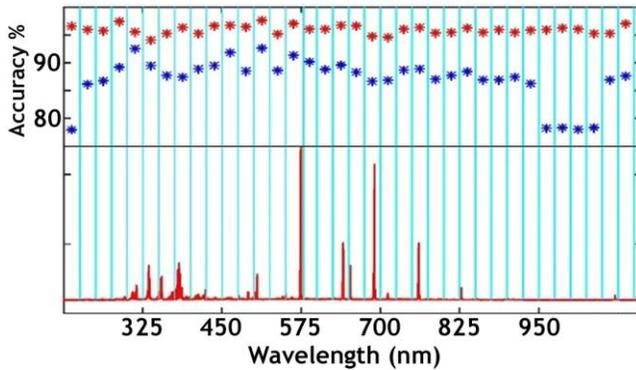

Figure 6: (Upper) Subset model accuracy for 36 subsets with ~100 wavelength variables each. Vertical lines indicate subset boundaries. Accuracy values are given for training samples (red) and test samples (blue). (Lower) Original spectrum example.

However this level of accuracy was not maintained when train and test followed protocol 4. The accuracy was found to fall considerably for all subsets with a maximum accuracy of 60% observed, Figure 7.

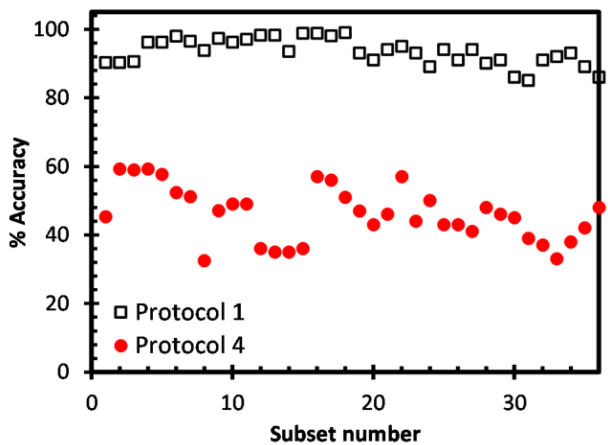

Figure 7: Subset model accuracy for 36 subsets with ~100 wavelength variables each. Comparison between Protocol 1 and Protocol 4.

Full wavelength variable models were further analysed by calculating the VIP scores to determine the relative contribution of each variable. For the simplest case of pure He, the scores versus wavelength plot, Figure 8, shows the highest VIP scores occur at spectral peaks. However the rank of the scores does not match the peak intensity rank, Table 4 i.e. the intensities of spectral peaks are not necessarily a good indicator of value to the model. Note, the VIP wavelengths do not exactly match the original peak values as the latter were rounded to indicate spectral variability. Selected VIPs are therefore associated with their nearest wavelength peak reported in Table 1. In the case of VIP rank 6 (391.7 nm), this is associated with original peak labelled 389 nm, which is a broad peak likely reflecting a main He line and other, much smaller, impurity features (CN, $N_2$, $O_2$)

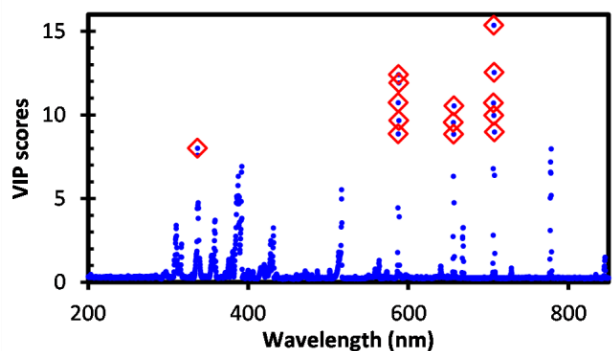

**Figure 8: VIP scores for whole spectra in pure He. The red diamonds highlight 14 VIPs with score ≥ 8. Each VIP corresponds to a single wavelength value.**

| VIP Rank | VIP wavelength | VIP score | Peak Intensity Rank | Species |
|---|---|---|---|---|
| 1 | 706.8 | 15.36 | 2 | He |
| 2 | 587.7 | 12.41 | 1 | He |
| 3 | 656.8 | 10.55 | 8 | H |
| 4 | 336.6 | 8.00 | 6 | Impurity |
| 5 | 778.2 | 7.97 | 4 | Impurity |
| 6 | 391.7 | 6.92 | 5 | He, CN, $N_2$, $O_2$ |
| 7 | 516.5 | 5.54 | >9 | C2 Swan |
| 8 | 309.9 | 3.41 | >9 | OH |
| 9 | 431.4 | 3.25 | >9 | CH |
| 10 | 668.1 | 3.24 | 3 | Impurity, He |

**Table 4: Comparison of VIP scores rank with peak intensity rank for pure He spectra.**

To create VIP feature selected models, an arbitrary VIP score threshold was chosen to balance the need for a manageable number of scores with the probability of including those most appropriate. For a threshold value of $N_{VIP} > 8$, 14 spectral peaks are selected in four wavelength regions, namely (i) 336.61 nm, (ii) 587.18 - 588.21 nm, (iii) 656.30 - 656.82 nm and (iv) 706.31 - 707.33 nm. A set of 14 reduced feature count PLS-DA models (9 LVs, Protocol 1) was created using ± 10 wavelength variables around each of the 14 VIP-selected peaks and their accuracy compared in Figure 9 (lower).

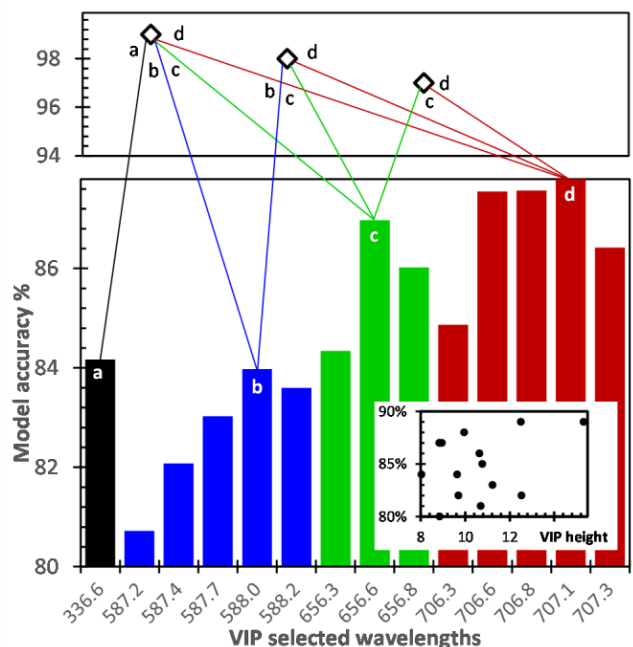

**Figure 9** *Lower*: Accuracy of 14 reduced feature count PLS-DA models, each based on ± 10 wavelength variables around single spectral peak centred at wavelengths where VIP scores > 8. Models were trained and tested according to Protocol 1 with 9 LVs. *Upper*: Accuracy of 3 reduced feature count PLS-DA models, each based on ± 10 wavelength variables around 2, 3 or 4 spectral peaks centred at wavelengths where VIP scores > 8. Models were trained and tested according to Protocol 1 with 9 LVs. The two-peak model (peaks c, d) uses VIP determined peaks at 707.07 nm and 656.56 nm, while 3 and 4 peak models use peaks (b, c and d) and (a, b, c and d) respectively. *Inset:* Model accuracy versus VIP height.

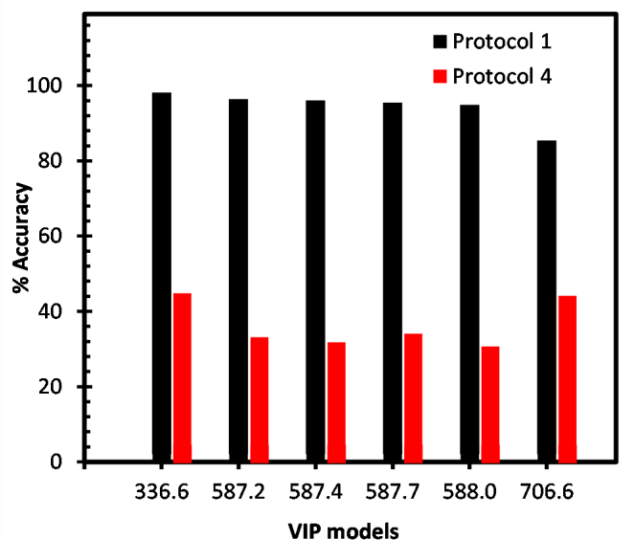

Figure 10: Accuracy of 6 reduced feature count PLS-DA models, each based on ± 10 wavelength variables around single spectral peaks centred at wavelengths (336.6 nm, 587.2 nm, 587.4 nm, 587.7 nm, 588.0 nm, 706.6 nm) where VIP scores > 8. Models were trained and tested according to Protocol 4 with 9 LVs and compared with Protocol 1.

With VIP feature selection, an accuracy of 88% is achievable using only 20 wavelength variables over a very restricted wavelength range. This compares to 92% accuracy achieved for the full model comprised of >3600 wavelengths. While there appears to be no relationship between VIP height and resultant model accuracy, Figure 9 (inset), a trend of increasing accuracy is apparent for sets containing the higher VIP scores, Figure 9 (lower). To further reduce the number of features, we selected the peaks corresponding to the highest scores from each of four sets and built PLS-DA models (2, 3 or 4 peaks, ± 10 wavelength variables per peak with 9 LVs). This resulted in an increase in accuracy to 99%, Figure 9 (upper). A further set of models were created for a number of VIP selected peaks (9 LVs, ±10 wavelength variables) with the training and testing carried out via Protocol 4. As occurred with data segmentation subset models, the accuracy fell significantly, with a maximum of 45%, Figure 10 .

While regularisation, data segmentation and VIP-related models reduced the wavelength variable count considerably (to 275, 100, and 20 respectively), only a limited number of variables are removed when using a peak width compression approach. After Savitzky-Golay smoothing, all spectral peaks above an arbitrary threshold height (100) are selected for compression, Figure 11. Overall, the total number of wavelength variables is reduced from the original 3668 depending on the extent of peak broadening e.g. for 6 ppm 3328 variables remain and 3300 for 100 ppm.

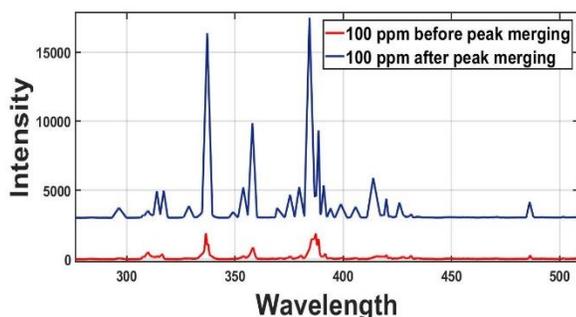

**Figure 11: Spectrum of He – CH$_4$ (100 ppm) sample before (red) and after (blue) peak compression in the wavelength interval 270 nm – 500 nm.**

The outcome of peak compression showed slight improvement in accuracy over results for protocols 1 and 2, Figure 4, but for protocol 4 a significant increase in accuracy was observed, Figure 12, reaching ≥ 97% for 8 LVs.

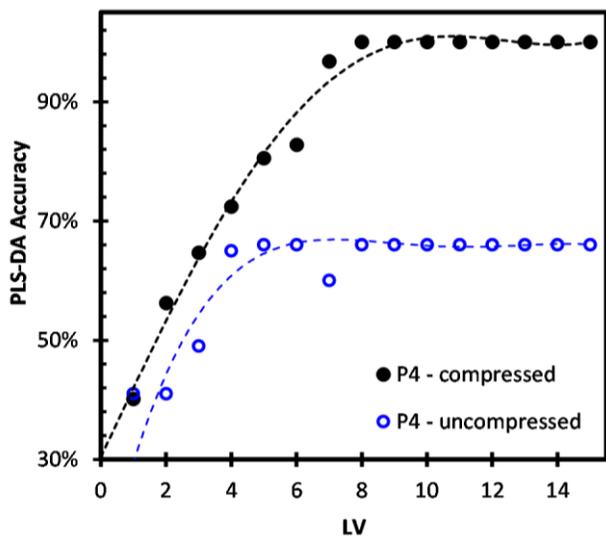

**Figure 12: PLS-DA accuracy versus LV using peak compression and trained / tested under Protocol 4, in comparison with accuracy obtained from uncompressed peaks.**

## 5. Discussion

Using PLS-DA classification applied to low resolution UV – visible range optical emission spectra derived from plasma excitation, we have demonstrated the ability to detect the presence of methane down to concentrations of 1 ppm and to label sample concentrations up to 100 ppm. Simple application of PLS-DA in protocols 1 and 2, with limited pre-processing, shows the capability of this algorithmic approach in developing accurate multi-categorical models based on high dimensionality spectral data. As expected, the accuracy increases with increasing number of latent variables used, levelling off in accuracy (> 90%) for LV ≥ 10. However the potential for overfitting of spectral data is obvious from protocol 3 where the mixed session data shows a fall in accuracy, for a given LV, compared to protocols 1 and 2. In protocol 4, the training and test data were from entirely different sessions presenting a much more realistic and stringent challenge which the algorithm failed to handle satisfactorily. These relatively simple plasma devices along with portable spectrometers inevitably produce highly variable output. Within a single session, RSD values for pure He can be > 10%, which increases to ~30% with added $CH_4$. Machine learning algorithms and associated pre-processing or enhancements offer the opportunity to directly negate the effect of this variability on predictive accuracy. They also offer the opportunity to gain further insight into underlying mechanisms to help improve issues such a plasma hardware, operating procedures and auto-filtering of data to allow progress to more complex sensing scenarios.

The data challenges faced in this work highlight one of the main practical difficulties with high dimensionality data. With a relatively small number of samples, models overfit to the training data and have reduced generality. The standard regularization approaches used here were unable to overcome the overfitting issue directly and appeared to penalize the data to the extent that featureless regions of the spectrum became primary predictors in the model. Our data segmentation approached showed that by reducing the number of predictor (wavelength) variables from ~3600 to 100 resulted in limited loss in model accuracy for protocols 1 and 2. This was observed for a number of subset models across the wavelength range 300 nm to 600 nm. However, even with such a reduced variable number, overfitting is still a significant factor and application of the approach using protocol 4 was unsuccessful. Nevertheless, reduction of variable count has been shown to be important not only for data analysis but may also allow use of lower specification and narrower range spectrometers, with implications for reduced cost. There is considerable scope in exploring the data segmentation approach further using multiple segment models and variable window sizes. Using VIP scores to reduce the variable count further, with models containing only 20 variables at single peaks, also resulted in high prediction accuracy for protocol 1 but much lower predictive success for protocol 4. This again indicates the prevalence of overfitting even though the number of wavelength variables is small compared to the number of samples. In contrast, the peak compression procedure has provided the greatest predictive success with regard to protocol 4 with accuracy values > 97%, an outcome as good as that obtained from Protocol 1. By reducing the multiple correlated wavelength variables around each spectral peak to a single variable, the effect of overfitting has been minimized. Also, since the compressed peak wavelengths are the same for each sample, the issue of spectrum misalignment is no longer a concern. Including segmentation and/or VIP selection with peak compression offers routes for consideration with more complex gas mixtures.

VIP score calculation is a technique by which the PLS-DA can report the significance of each individual variable (wavelength) to the model predictions and as such provides direct physical insight into the primary plasma factors underlying the model. By definition, the average VIP score is one and is often used as a significance threshold. However, in order to restrict the wavelength priority list to a manageable number we used a high threshold value (VIP > 8), producing a list of 14 wavelengths. Analysis of the full variable (>3600) count models using VIP indicated, as expected, that the primary contributors to models were located at the spectral peaks. However, the spectral peak height ranking did not necessarily follow the ranking of the VIP scores. Within this list of top 10 VIP scores, the top two represent the highest peaks from He emission, while of the remainder, five can be attributed to hydrogen, nitrogen or oxygen related impurities, with relatively small peaks, and two to carbon-based species.

The nature and impact of chemical species on the algorithm prediction is important on two counts. If the algorithm were to be dependent solely on He peaks, then the applicability of the technique with other plasma gases, e.g. Ar, $N_2$ or air, is unclear. A dependence on hydrocarbon impurity peaks may imply a high degree of $CH_4$ dissociation which could hamper attempts to differentiate different hydrocarbons. With the discovery of non-$CH_4$ impurity emission as a significant factor in the VIP score list, it is possible however that prediction depends on $CH_4$ – induced changes to the overall plasma. As is observed with molecular gases in general, we would expect collisions with low-energy electrons to result in vibrational excitation of the molecule while the absorption of these electrons may lead to change in the EEDF sufficient to affect the emission of all species. The development of suitable plasma chemistry models is severely hampered by the limited rate coefficient and cross-section data for many of the possible reactions and the lack of experimental plasma parameter values. Nevertheless, it is worthwhile assessing the potential significance of both $CH_4$ dissociation and non-$CH_4$ impurity impact on EEDF.

Given the likely complexity of the plasma chemistry along with the limited spectrometer resolution, multiple species assignment to a single emission line is possible and while knowledge of the underlying chemistry would be valuable, it is not currently available. In Vincent et al., we discuss the possible He – $CH_4$ chemistry at trace methane levels and its impact on emission spectra.[35] Molecular $CH_4$ has no emission lines in the wavelength range 200 nm – 1100 nm. However we observe small features around 431 nm and 389 nm which can be attributed to CH emissions from the $A^2\Delta \rightarrow X^2\Pi$ system [65] and the (0,0) band of the $B\ ^2\Sigma^- \rightarrow X\ ^2\Pi$ system, although the 389 nm peak also represents emission from the He transition (1s2s - 1s3p). [48] These features are weak and the variance is relatively large, nevertheless there is a trend of increasing intensity with $CH_4$ concentration. The H$\alpha$ line at 656.28 nm is present in all spectra but becomes the dominant peak at $CH_4$ concentrations above ~40 ppm. At low $CH_4$ concentrations the dissociation of $H_2O$ may be the primary source of H$\alpha$; observed peaks around 310 nm are likely due to OH(A-X) emission.[33] From high resolution humidity measurements of the plasma source gas, we estimate $H_2O$ content between 10 ppm and 500 ppm in our pure He plasmas. Emission due to $C_2$ Swan vibrational bands around 516 nm appears at concentrations above 77 ppm. These correspond to transitions between the $d^3\Pi_g$ (2.48 eV) and $a^3\Pi_u$ (0.09 eV) electronic states and indicate the final hydrogen abstraction endpoint from $CH_4$. CH emission normally dominates over $C_2$ emission in hot methane flames or plasmas [66] since the latter derives from $C_2H_y$ species which in turn are produced by heavy particle collisions between methane radicals, e.g. dimerisation reactions between CH and $CH_x$. [66] These reactions are often exponentially dependent on gas temperature, with thresholds typically >

1000 °C.[65] Therefore $CH_4$ dissociation and emission from $C_xH_y$ fragments (x: 0 → 2, y: 0 → 4) can be expected to make some contribution to PLS-DA models.

The presence of $H_2O$, $N_2$ and $O_2$ impurities and their associated radicals also lead to additional emission features. For example, a persistent peak around 336 nm can be attributed to $N_2$ rotational and vibrational molecular bands. To estimate the effect of $CH_4$ or $CH_4$ plus some dissociation fragments on pure He emission with or without molecular impurities, we calculated the electron energy distribution functions (EEDF) of different mixtures using a Boltzmann solver [67] along with the available cross-sections for $CH_4$, some related $CH_4$ dissociation reactions [68] as well as those for $H_2O$, $N_2$ and $O_2$.[69] Calculated rate equations for $CH_4$ elastic collisions are similar to those of helium and given the trace level $CH_4$ concentrations, the calculated electron energy loss due to He elastic collisions remains unchanged on the introduction of trace gases. We observed no change in absorbed power as $CH_4$ is added to the helium plasma. InThe analysis of EEDF variation due to the inclusion of trace impurities and hydrocarbons has limited direct predictive capability at this stage due to a lack of information on plasma chemistry at atmospheric pressure and cross-section details for a large number of potential reactions. Nevertheless, even with small quantities of impurities or hydrocarbons, the change in EEDF is noticeable. This is expected since a high rate of vibrational/rotational excitation occurs in molecular gases as well as dissociation, at energies well below those of pure noble gases. Overall, with the objective being trace gas detection, we observe spectral changes due to additional impurity and hydrocarbon peaks as well as changes to the primary He peaks, due to impurity and hydrocarbon induced EEDF modification. According to the priority VIP list, the latter is a significant factor in the algorithm operation.

Figure 13 (a), the impact of $CH_4$ (1 ppm) and impurities ($H_2O$ 500 ppm, $O_2$ 10 ppm, $N_2$ 10 ppm, and H 10 ppm) on the pure He EEDF can be observed. The addition of $CH_4$ tends to increase the high energy tail of the EEDF with this effect decreasing as the concentration reaches 100 ppm, The analysis of EEDF variation due to the inclusion of trace impurities and hydrocarbons has limited direct predictive capability at this stage due to a lack of information on plasma chemistry at atmospheric pressure and cross-section details for a large number of potential reactions. Nevertheless, even with small quantities of impurities or hydrocarbons, the change in EEDF is noticeable. This is expected since a high rate of vibrational/rotational excitation occurs in molecular gases as well as dissociation, at energies well below those of pure noble gases. Overall, with the objective being trace gas detection, we observe spectral changes due to additional impurity and hydrocarbon peaks as well as changes to the primary He peaks, due to impurity and hydrocarbon induced EEDF modification. According to the priority VIP list, the latter is a significant factor in the algorithm operation.

Figure 13 (b). However, atmospheric impurity species have almost the opposite effect of decreasing the higher energy regions of the EEDF. Nevertheless, the addition of $CH_4$ (1 ppm) to He with impurities included tends to negate this effect, The analysis of EEDF variation due to the inclusion of trace impurities and hydrocarbons has limited direct predictive capability at this stage due to a lack of information on plasma chemistry at atmospheric pressure and cross-section details for a large number of potential reactions. Nevertheless, even with small quantities of impurities or hydrocarbons, the change in EEDF is noticeable. This is expected since a high rate of vibrational/rotational excitation occurs in molecular gases as well as dissociation, at energies well below those of pure noble gases. Overall, with the objective being trace gas detection, we observe spectral changes due to additional impurity and hydrocarbon peaks as well as changes to the primary He peaks, due to impurity and hydrocarbon induced EEDF modification. According to the priority VIP list, the latter is a significant factor in the algorithm operation.

Figure 13 (c), and the high energy tail increases. The impact of $CH_x$ dissociation species on EEDF also shows a complex relationship with concentration. While EEDFs for He – $CH_4$ (100 ppm) and He – $CH_4$ (0 ppm) are almost indistinguishable, the presence of 10 ppm $CH_X$ in He – $CH_4$ (100 ppm) leads to a significant decrease in EEDF between mean electron energies of 4 – 13 eV before rising again at higher energies, The analysis of EEDF variation due to the inclusion of trace impurities and hydrocarbons has limited direct predictive capability at this stage due to a lack of information on plasma chemistry at atmospheric pressure and cross-section details for a large number of potential reactions. Nevertheless, even with small quantities of impurities or hydrocarbons, the change in EEDF is noticeable. This is expected since a high rate of vibrational/rotational excitation occurs in molecular gases as well as dissociation, at energies well below those of pure noble gases. Overall, with the objective being trace gas detection, we observe spectral changes due to additional impurity and hydrocarbon peaks as well as changes to the primary He peaks, due to impurity and hydrocarbon induced EEDF modification. According to the priority VIP list, the latter is a significant factor in the algorithm operation.

Figure 13 (d). For example, the impact of this change in EEDF on the He emission (1s3d→1s2p, 587.56 nm) is illustrated by the variation in calculated rate coefficients for the He 1s3d excitation, The analysis of EEDF variation due to the inclusion of trace impurities and hydrocarbons has limited direct predictive capability at this stage due to a lack of information on plasma chemistry at atmospheric pressure and cross-section details for a large number of potential reactions. Nevertheless, even with small quantities of impurities or hydrocarbons, the change in EEDF is noticeable. This is expected since a high rate of vibrational/rotational excitation occurs in molecular gases as well as dissociation, at energies well below those of pure noble gases. Overall, with the objective being trace gas detection, we observe spectral changes due to additional impurity and hydrocarbon peaks as well as changes to the primary He peaks, due to impurity and hydrocarbon induced EEDF modification. According to the priority VIP list, the latter is a significant factor in the algorithm operation.

Figure 13 (e). The rate coefficient is very sensitive to the mean electron energy ($\varepsilon$), falling sharply around 2 eV before increasing up to 10 eV. Using an expected value of $\varepsilon$ = 2 eV, the calculated reduction in rate coefficient is ~32%. Experimentally the observed reduction in the He emission (1s3d→1s2p, 587.56 nm) peak is 35% - 40% with the addition of 100 ppm $CH_4$. The analysis of EEDF variation due to the inclusion of trace impurities and hydrocarbons has limited direct predictive capability at this stage due to a lack of information on plasma chemistry at atmospheric pressure and cross-section details for a large number of potential reactions. Nevertheless, even with small quantities of impurities or hydrocarbons, the change in EEDF is noticeable. This is expected since a high rate of vibrational/rotational excitation occurs in molecular gases as well as dissociation, at energies well below those of pure noble gases. Overall, with the objective being trace gas detection, we observe spectral changes due to additional impurity and hydrocarbon peaks as well as changes to the primary He peaks, due to impurity and hydrocarbon induced EEDF modification. According to the priority VIP list, the latter is a significant factor in the algorithm operation.

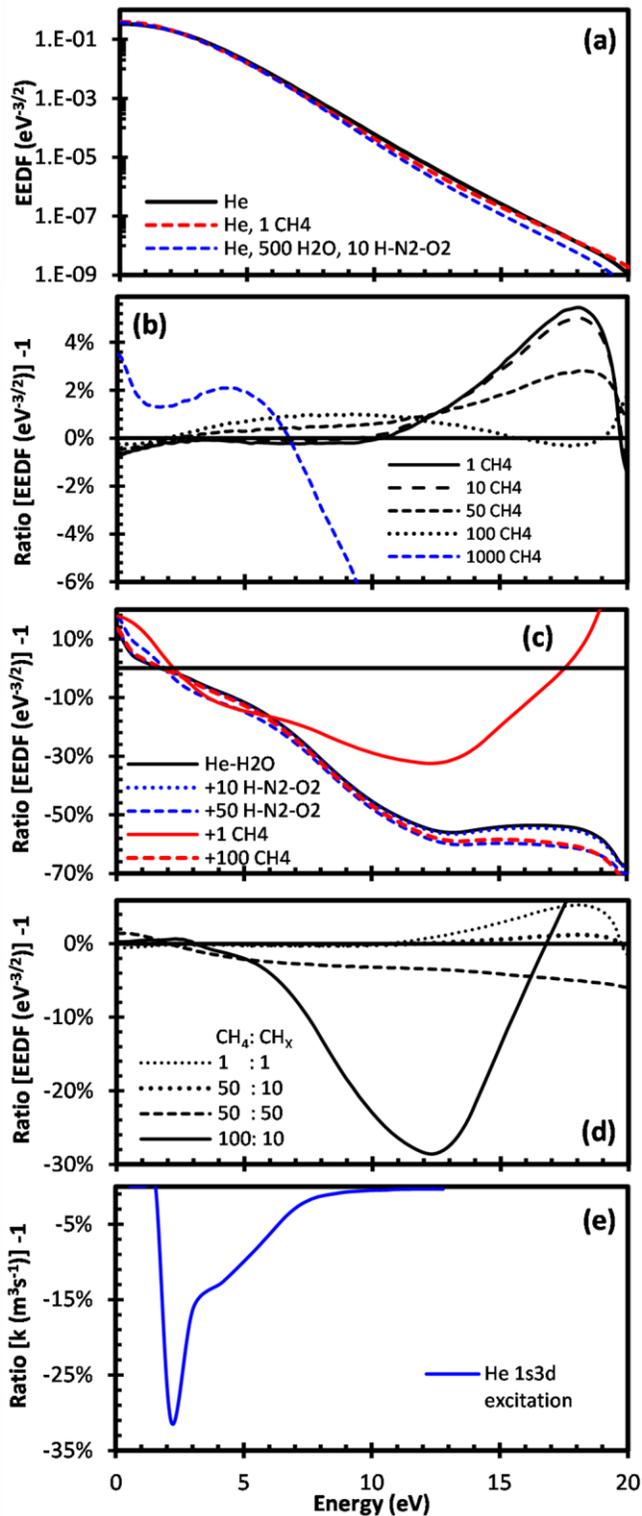

Figure 13: (a) EEDF plots for He, He + CH$_4$ (1 ppm), He + H$_2$O (500 ppm) + N$_2$/O$_2$ impurities (10 ppm), (b) ratio of EEDF vs energy with addition of CH$_4$ to the EEDF of He (0 ppm), (c) change in impurity to helium EEDF ratio with the addition H$_2$O (500 ppm) to He (black), with addition of H, N$_2$ and O$_2$ impurities to He – H$_2$O (500 ppm) (blue), and with the addition of CH$_4$ to He – H$_2$O/H/N$_2$/O$_2$ (red), (d) change in EEDF ratio with the addition of CH$_4$ and CH$_x$ (x: 0→3) to He, (e) variation

in the ratio of rate coefficients for the He 1s3d excitation in He and in He + $CH_4$/$CH_x$ (100 ppm / 10 ppm).

6. Conclusion

We have demonstrated the capability of using optical emission spectroscopy from a small-volume (5 μL) atmospheric pressure plasma, coupled with PLS-DA spectral classification algorithms, to detect the presence of methane down to concentrations of 1 ppm and to label sample concentrations up to 100 ppm. This compares well with portable NDIR systems [10], which deliver LOD values above 50 ppm, and low cost chemi-resistive sensors which represent the most commonly deployed technology.[70] The ability to detect $CH_4$ and assign a concentration classification offers scope for higher resolution classifications which will be valuable for diagnostics, online monitoring of trends and developing advanced warning capabilities. Nevertheless, future plasma emission sensor devices will also need to handle increased levels of matrix gases including air and other hydrocarbons and will require further development of algorithms and plasma sources. We have investigated a number of algorithm enhancements including regularization, simple data segmentation and subset selection, VIP feature selection and wavelength variable compression. All these approaches showed the potential for significant reduction in the number of wavelength variables and the spectral resolution/bandwidth – an important technological consideration. However only wavelength variable compression exhibited reliable predictive performance under the more challenging multi-session train – test scenarios. Nevertheless, there is still considerable scope for fine tuning the application of single and multiple enhancements. Gaining some understanding of plasma – gas interactions, their appearance in spectra and their interpretation by classification algorithms is important for algorithm enhancement when faced with a wide array of options. Although knowledge of radical species densities and their cross-sections is very limited, modelling the impact of chemistry on plasma conditions has illustrated the complex cross-sensitivities in the excitation of noble gas, impurities, target $CH_4$ and its dissociation fractions. The discovery that trace impurity species variation, other than in the target gas, is a significant factor in algorithm prediction indicates that successful operation can be possible independent of the choice of plasma gas.